\documentclass[12pt]{article}
\usepackage{a4wide,amssymb,cite}
\pdfoutput=1

\usepackage{a4wide,amssymb,graphicx}
\usepackage{epsfig}
\usepackage[usenames,dvipsnames]{color}
\usepackage{slashed}
\newcommand{\be}{\begin{equation}}
\newcommand{\ee}{\end{equation}}
\newcommand{\bea}{\begin{eqnarray}}
\newcommand{\eea}{\end{eqnarray}}

\def\circa#1{\,\raise.3ex\hbox{$#1$\kern-.75em\lower1ex\hbox{$\sim$}}\,}

\begin{document}

\begin{titlepage}
%
%

%\rightline{CERN-PH-TH/2014-025}

%

\begin{centering}
\vspace{1cm}
{\Large {\bf Gauged $U(1)$ Clockwork Theory}} \\

\vspace{1.5cm}

{\bf  Hyun Min Lee}
%\\
\vspace{.5cm}

{\it Department of Physics, Chung-Ang University, Seoul 06974, Korea.} 
\\

\end{centering}
\vspace{2cm}

\begin{abstract}
\noindent
We consider the gauged $U(1)$ clockwork theory with a product of multiple gauge groups and discuss the continuum limit of the theory to a massless gauged $U(1)$ with linear dilaton background in five dimensions. The localization of the lightest state of gauge fields on a site in the theory space naturally leads to exponentially small effective couplings of external matter fields localized away from the site. 
We discuss the implications of our general discussion with some examples, such as mediators of dark matter interactions, flavor-changing $B$-meson decays as well as D-term SUSY breaking.

\end{abstract}

\vspace{3cm}

\end{titlepage}

\section{Introduction}

 A large field excursion of the axion-like field in the effective theory can be realized in the setup with multiple fields due to a sequential suppression of the effective axion couplings \cite{aligned,choi,rattazzi,axion}, which is the so called the clockwork theory in a general term. The clockwork theory has been recently proposed as a general framework where small or large effective couplings  to the lightest state at low energy can be naturally obtained in the presence of multiple fields with nearest neighbored interactions \cite{gian1,gian2,craig}. There have been some applications of the clockwork theory to inflation \cite{inflation}, dark matter \cite{DM}, and other interesting particle physics problems \cite{etc}. 
 
The concrete realization of clockwork theory requires a product of multiple identical symmetry groups and fields (or clock gears) with asymmetric charges under the neighbored symmetries. Since the local non-abelian symmetries lead to charge quantization condition, the clockwork mechanism for local symmetries is restricted only to the case with abelian symmetries, such as local $U(1)$ symmetries \cite{craig,batell2}. 

In this article, we consider a gauged $U(1)$ clockwork, which leads to the zero mode of gauge fields with a position-dependent coupling to the matter fields localized on the sites \cite{gian1,craig,gian2}. On the other hand, the massive modes of gauge fields can have sizable couplings to the localized matter fields, so they could be accessible in the collider experiments. 
For instance, the zero mode of the $U(1)$ clockwork may play a role of light mediator for dark matter which is localized at the peak of the zero mode wave function, while evading the bound from direct and indirect detections of dark matter if the SM particles are localized at another position in the tail of the zero mode of the $U(1)$ clockwork.  

We discuss the construction of the gauged $U(1)$ clockwork theory and take the continuum limit of the theory in five dimensions. 
Then, we introduce localized interaction terms of external fields to $U(1)$ clock gears and illustrate some concrete examples for utilizing the gauged $U(1)$ clockwork  such as mediators of dark matter interactions, $B$-meson decays, and D-term supersymmetry (SUSY) breaking.
Finally, conclusions are drawn.

\section{The setup}

We consider a  local $U(1)$ clockwork, composed of $N+1$ independent $U(1)$ gauge fields, $A^j_\mu (j=0,1,\cdots,N)$, and a set of $N$ Higgs fields, $\phi_j(j=0,1,\cdots,N-1)$. For a continuum limit, we also need to add a set of $N-1$ scalar fields, $S_j(j=0,1,\cdots,N-2)$, linking the Higgs fields \cite{pokorski}. The $U(1)$ charges are assigned asymmetrically between two neighbored $U(1)$'s as $\phi_0=(1,-q,0,0,\cdots,0)$, $\phi_1=(0,1,-q,0,0,\cdots, 0)$, $\cdots$, $\phi_{N-1}=(0,0,\cdots, 0,1,-q)$, and $S_0=(1,-q-1,q,0,\cdots)$, $S_1=(0,1,-q-1,q,0,\cdots)$, $\cdots$, $S_{N-2}=(0,0,\cdots,1,-q-1,q)$.  After the Higgs fields $\phi_j$ get VEVs, the product group of $N+1$ U(1)'s, $U(1)_0\times U(1)_1\times\cdots\times U(1)_{N+1}$, is broken down to one $U(1)$.

The Lagrangian for the gauged $U(1)$ clockwork is given by
\bea
{\cal L}&=&-\sum_{j=0}^N \frac{1}{4} F^j_{\mu\nu} F^{j\mu\nu} - \sum_{j=0}^{N-1} \bigg[(D_\mu \phi_j)^\dagger D^\mu\phi_j+ V(\phi_j) \bigg]  + \sum_{j=0}^{N-2} m^2_0 \Big| \phi_j-\omega^{-1}\,S_j\phi_{j+1}\Big|^2 \nonumber \\
&=&-\sum_{j=0}^N \frac{1}{4} F^j_{\mu\nu} F^{j\mu\nu} - \sum_{j=0}^{N-1} \bigg[(D_\mu \phi_j)^\dagger D^\mu\phi_j  +2 m^2_0 |\phi_j|^2+ V(\phi_j) \bigg]-\sum_{j=0}^{N-2}(\alpha  \phi^\dagger_j S_j\phi_{j+1}+{\rm h.c.} )\nonumber \\
&&\quad +m^2_0 |\phi_0|^2 +m^2_0 |\phi_{N-1}|^2  \label{Lag}
\eea
where $\alpha\equiv \frac{m^2_0}{\omega}$, the covariant derivative is defined as $D_\mu\phi_j=(\partial_\mu+ig(A^j_\mu-q A^{j+1}_\mu))\phi_j$,  and the Higgs potential at each site is given by
\be
V(\phi_j)= -{\tilde m}^2 |\phi_j|^2+ {\tilde\lambda} |\phi_j|^4,
\ee
and the VEV of the link fields $S_j$ are taken such that the common mass terms proportional to $m^2_0$ are written and the quartic interactions to the link fields $S_j$ are omitted as in the second line. 

The Higgs potential $V(\phi_j)$ leads to the Higgs VEV at each site, $\langle\phi_j\rangle=\frac{1}{\sqrt{2}}\, v_j$, which are set to be universal as $v_j=f (j=0,1,\cdots, N-1)$. 
This VEV choice leads to $\langle S_j\rangle=\omega$ for generating no tadpole from the mass terms linking different Higgs fields.  We assume that $\omega\ll f$ such that the additional mass terms for gauge fields due to $\langle S_j\rangle$ are negligible. 
Then, expanding the Higgs fields around the VEV as $\phi_j=\frac{1}{\sqrt{2}}(f+h_j) e^{i\pi_j/f}$ and taking $S_j=\omega $ with ignoring the fluctuations for $S_j$, the Higgs part of the above Lagrangian becomes
\bea
{\cal L}_{\rm Higgs}&=& -\sum_{j=0}^{N-1}\bigg(\frac{1}{2}(\partial_\mu h_j)^2 + \frac{1}{2} m^2_0 |h_j-h_{j+1}|^2  \nonumber \\
&&\quad + \frac{1}{2}g^2 (f+ h_j)^2 \Big(A^j_{\mu}-q A^{j+1}_\mu+\frac{1}{gf} \partial_\mu\pi_j\Big)^2+V(f+h_j) \bigg).  \label{Higgs}
\eea
Here, we note that the gauge invariant mass terms linking between $\phi_j$ and $\phi_{j+1}$ do not produce the mass terms for the would-be Goldstone bosons, $\phi_j$, after the $U(1)$'s are broken spontaneously.  

We remark that the Higgs boson interactions would not be necessary for a Stueckelberg formulation for massive gauge bosons in the gauged $U(1)$ clockwork theory.
But, we have kept the Higgs boson interactions explicitly for renormalizability in the discrete clockwork theory. As a result, the Higgs kinetic term along the extra dimension is generated from the Higgs mass terms in the continuum limit, as will be discussed in the next section. We will also show later that the Higgs bosons are decoupled from the rest of the fields in the continuum limit.

\subsection{Gauge fields}

We first discuss the mass spectrum and mass eigenstates of gauge fields. 
Introducing the appropriate gauge fixing term to cancel the mixing terms between gauge fields and would-be Goldstone bosons, the gauge boson mass terms are written as
\bea
{\cal L}_{\rm gauge}= -\sum_{j=0}^{N-1}\frac{1}{2}g^2 f^2  \Big(A^j_{\mu}-q A^{j+1}_\mu\Big)^2.
\eea
Then, the massless mode of gauge field is given by
\be
{\tilde A}^0_\mu(x)=  \sum_{j=0}^N a_{j0}\, A^j_\mu(x),\quad  \label{massless}
\ee
with 
\be
a_{j0}=\frac{N_0}{q^j}, \quad N_0= \sqrt{\frac{q^2-1}{q^2 -q^{-2N}}}.
\ee
On the other hand, we obtain the massive modes of gauge field as follows,
\bea
{\tilde A}^k_{\mu}(x)= \sum_{j=0}^N a_{jk} \,A^j_\mu(x)  \label{massive}
\eea
with the mass eigenvalues given by
\be
M^2_k= m^2\Big(1+q^2-2q \cos \frac{k\pi}{N+1} \Big)\equiv m^2\lambda_k, \quad m^2\equiv g^2 f^2, \quad k=1,2,\cdots, N,
\ee
where the wave functions are given by
\bea
a_{jk} = N_k\left[q\sin \Big(\frac{jk\pi}{N+1}\Big)-\sin \Big(\frac{(j+1)k\pi}{N+1} \Big)\right] , \quad N_k=\sqrt{\frac{2}{(N+1)\lambda_k}}.
\eea
We can easily obtain the interacting gauge fields by inverting eqs.~(\ref{massless}) and (\ref{massive}), as follows,
\bea
{A}^j_{\mu}(x)= \sum_{k=0}^N a_{jk} \,{\tilde A}^j_\mu(x), \quad j=0,1,2,\cdots, N. \label{interaction} 
\eea
For instance, the gauge field at site $l$ is expanded in terms of mass eigenstates as follows,
\bea
A^l_\mu(x)= \frac{N_0}{q^l}\, {\tilde A}^0_\mu(x)+\cdots.
\eea

\subsection{Higgs fields}

Ignoring the Higgs potential at each site, the Higgs mass terms are given by
\bea
{\cal L}_{\rm Higgs}= -\sum_{j=0}^{N-1}\frac{1}{2} m^2_0 |h_j-h_{j+1}|^2. 
\eea
Therefore, the mass eigenvalues and wave functions can be similarly obtained. 
But, since the nearest neighbor interactions are symmetric in this case, i.e. $q=1$, the wave function of the Higgs zero mode is massless and flat whereas the mass eigenvalues of massive modes are given by
\bea
M^2_k= 4m^2_0 \sin^2\Big(\frac{k\pi}{N}\Big), \quad k=1,2,\cdots, N-1. 
\eea
In the presence of the common mass terms for Higgs fields at each site, the mass spectrum of zero mode and massive modes are modified to
\bea
M^2_k=4m^2_0 \sin^2\Big(\frac{k\pi}{N}\Big)+ {\tilde m}^2_h, \quad  k=1,2,\cdots, N-1.
\eea
where ${\tilde m}^2_h=2 {\tilde\lambda} f^2$.

\section{The continuum limit}

In this section, we discuss the continuum limit of the gauged $U(1)$ clockwork for both gauge and Higgs fields and identify the corresponding Lagrangian in a five-dimensional field theory with linear dilaton background.

\subsection{Bulk gauge field}

Similarly as in deconstructing dimensions with symmetric interactions \cite{AH,pokorski0,pokorski}, we can take the continuum limit of the discrete $U(1)$ clockwork by introducing a lattice distance $a$ with $a\rightarrow 0$ and $f\rightarrow \infty$ while keeping $af$ finite and $a g f q=1$ and $q\rightarrow 1$.  Introducing the extra coordinate as $y=j a$ such that $y=0$ for $j=0$ and $y=\pi R\equiv N a$, we identify $A_\mu (x,y)\equiv A^j_\mu(x)$ and the would-be Goldstone bosons beomces the extra component of gauge field by $A_y(x,y)\equiv \pi_j(x)$.

Then, the gauge part of the Lagrangian (\ref{Lag}) becomes in the continuous limit
\bea
{\cal L}_{\rm gauge}&=& \int^{\pi R}_0 dy \bigg[-\frac{1}{4} F_{\mu\nu} F^{\mu\nu}-\frac{1}{2} \Big(\partial_y A_\mu-\partial_\mu A_y+k A_\mu\Big)^2 \bigg] \nonumber \\
&=&  \int^{\pi R}_0 dy \bigg[ -\frac{1}{4} F_{MN} F^{MN}-\frac{1}{2}A^2_\mu \Big(k^2 -2k (\delta(y)-\delta(y-\pi R))\Big)+ k A_\mu \partial^\mu A_y \bigg]
\eea
where $k\equiv \frac{q-1}{qa}$ and the 5D indices are $M,N=0,1,2,3,5$. 
As a result, the localization of the zero mode is achieved at the expense of bulk and brane-localized mass terms for gauge bosons \cite{batell1,craig,gian2}. 

Choosing another field basis for gauge field by $B_M= e^{ky} A_M$, we can eliminate the gauge boson masses and rewrite the continuum gauge Lagrangian in a fully 5D Lorentz invariant form, as follows,
\bea
{\cal L}_{\rm gauge}= \int^{\pi R}_0 dy  \, e^{-2ky} \bigg[-\frac{1}{4} F_{MN} F^{MN}\bigg].
\eea
Therefore, the effective gauge coupling depends on the location in the extra dimension by the dilaton factor, $e^S=e^{-2ky}$, in the linear dilaton background on the $S^1/Z_2$ orbifold \cite{lindilaton,gian1}.
Then, as for the scalar clockwork \cite{gian1}, the equation of motion for the bulk $U(1)$ gauge field is similarly given by
\bea
\partial_M\Big(\sqrt{-g} \, e^S F^{MN}\Big)=0.   \label{5Deq}
\eea

Taking the flat 5D metric, $ds^2=g_{MN} dx^M dx^N=\eta_{\mu\nu} dx^\mu dx^\nu+ dy^2$, and the $A_y=0$ gauge (or unitary gauge), from eq.~(\ref{5Deq}), we derive  the equation for the 4D components of $U(1)$ with $A_\mu(x,y)=\sum_n {\tilde A}^{(n)}_\mu(x) f_n(y)/\sqrt{\pi R}$, becomes
\bea
f^{\prime\prime}_n - 2k f'_n +m^2_n f_n=0  \label{feq}
\eea
where prime denotes the derivative with respect to $y$ and $(\Box-m^2_n){\tilde A}^{(n)}_\mu=0$. 
Then, making the field redefinition with $f_n= e^{ky} \psi_n$, eq.~(\ref{feq}) becomes
\bea
\psi^{\prime\prime}_n - k^2 \psi'_n +m^2_n f_n=0. 
\eea
As a consequence, we obtain the wave function of the zero mode as  
\be
\psi_0= N_0 \, e^{-k|y|}  \label{zero}
\ee 
with normalization factor
\bea
N_0=\sqrt{\frac{k\pi R}{1-e^{-2k\pi R}}}.  \label{norm0}
\eea 
On the other hand, the massive eigenstates are determined to be
\bea
\psi_n(y)= N_n  \Big(\cos\frac{ny}{R} - \frac{kR}{n}\,\sin\frac{n|y|}{R} \Big)  \label{massive}
\eea 
where the mass eigenvalues and the normalization factor are, respectively, 
\bea
m^2_n&=&k^2+ \frac{n^2}{R^2},  \label{mass} \\
N_n&=&\frac{n}{m_n R}  \label{norm}
\eea
with $n=0,1,2\cdots$. 
Here, we took the modulus $|y|$ for mass eigenstates defined in the covering space $S^1$ and use was made of the normalization condition, $\int^{\pi R}_{-\pi R} \frac{dy}{\pi R}\,  \psi_n(y) \psi_m(y)=\delta_{nm}$.  Thus, $k$ leads to a mass gap between and the zero mode and the massive modes and $R\gg k^{-1}$ leads to the compact spectrum of massive modes.
Consequently, we find that the results are consistent with those in the discrete clockwork in the continuum limit.

\subsection{Bulk Higgs field}

On the other hand, taking the continuum limit for the Higgs sector by keeping $a, {\tilde\lambda} \rightarrow 0$ and $f, m_0\rightarrow \infty$ while $m_0 a=1$ and and ${\tilde\lambda} f^2$ finite, we also obtain the corresponding Lagrangian as 
\bea
{\cal L}_{\rm Higgs}= - \int^{\pi R}_0 dy \bigg[ \frac{1}{2}(\partial_M h)^2 + \frac{1}{2} {\tilde m}^2_h h^2 \bigg]
\eea
Here, we find that in the limit of $f\rightarrow \infty$ in our continuum limit, the gauge-Higgs interactions vanish in the Higgs part (\ref{Higgs}), so the gauge and Higgs sectors are completely decoupled. 
As a consequence, the gauge field requires a nontrivial coupling to the dilaton while the Higgs field does not.

\section{Interactions to $U(1)$ clock gears}

We consider the interactions of external matter fields to $U(1)$ clock gears and the breakdown of the remaining $U(1)$ (the zero mode of $U(1)$ clock gears) due to the Higgs mechanism localized on one site. The results in this section can be used for a later discussion on the examples for the $U(1)$ clockwork.

\subsection{Couplings of $U(1)$ clockwork to external fields}

In the discrete $U(1)$ clockwork, we can introduce the $U(1)$ interaction to an external matter field localized at one site. 

Suppose that a fermion $\psi$ with nonzero charge under $U(1)_l$ is introduced at site $l$ as follows,
\be
{\cal L}_{\rm fermion}= i {\bar\psi}\gamma^\mu\Big(  \partial_\mu + i g (v_\psi + a_\psi \gamma^5)A^l_\mu(x)\Big) \psi(x). 
\ee
Then, from the expansion of the gauge fields $A^l_\mu$ given in eq.~(\ref{interaction}), we obtain the couplings of the fermion to mass eigenstates of gauge fields as
\bea
{\cal L}_{\rm f, int}= - g {\bar\psi}(x)\gamma^\mu(v_\psi +a_\psi \gamma^5) \psi(x) \left(\frac{N_0}{q^l}\, {\tilde A}^0_\mu(x)+\sum_{k=1}^N  a_{lk} {\tilde A}^k_\mu(x) \right).  \label{fermion}
\eea
As a result, we can obtain the effective gauge coupling to be exponentially suppressed as $g_{\rm eff}=\frac{ N_0 g}{q^l}$, due to the localization of the zero mode of gauge field.

In the continuum limit, we can introduce a vector-like fermion $\psi$ localized at $y=y_0$ as follows, 
\be
{\cal L}'_{\rm fermion}= \int dy\,  \delta(y-y_0)\,  i{\bar \psi}(x) \gamma^\mu \Big(\partial_\mu + i g_5 (v_\psi + a_\psi \gamma^5)A_\mu(x,y)\Big) \psi(x)
\ee
where $g_5$ is the 5D gauge coupling.
Then, as the zero mode of gauge field has a profile, $A^0_\mu\sim e^{-k y}$, the effective gauge coupling to the localized fermion at $y=y_0$ can be exponentially suppressed as $e^{-k y_0}$.

On the other hand,  we consider an extra complex scalar field $\phi$ with charge $+1$ on the site $k$, with the following Lagrangian,
\bea
{\cal L}_{\rm scalar}= \Big|\partial_\mu \phi+i g A^l_\mu(x) \phi\Big|^2-V(\phi)
\eea
where the scalar potential for the extra scalar is given by $V(\phi)=-m^2_\phi |\phi|^2+\lambda_\phi |\phi|^4$. 
Then, the $U(1)$ interactions to the mass eigenstates of gauge fields are given by
\bea
{\cal L}_{\rm s, int}&=& -ig (\phi^*\partial^\mu\phi- \phi\partial^\mu\phi^*)A^l_\mu(x)+g^2 (A^l_\mu(x) )^2|\phi|^2 \nonumber \\
&=&-i g(\phi^*\partial^\mu\phi- \phi\partial^\mu\phi^*) \left(\frac{N_0}{q^l}\, {\tilde A}^0_\mu(x)+\sum_{k=1}^N  a_{lk} {\tilde A}^k_\mu(x) \right) \nonumber \\
&&+ g^2 \left(\frac{N_0}{q^l}\, {\tilde A}^0_\mu(x)+\sum_{k=1}^N  a_{lk} {\tilde A}^k_\mu(x) \right)^2 |\phi|^2. \label{scalar}
\eea
Therefore, we find that the remaining $U(1)$ invariance is manifest from the relation between linear and quadratic gauge interactions of the zero mode. 

Similarly to the fermion external fields, we can take the continuum limit by introducing a complex scalar field $\phi$ at $y=y_0$, leading to 
\bea
{\cal L}'_{\rm scalar}= \int dy\, \delta(y-y_0) \left( \Big|\partial_\mu \phi(x)+i g_5 A_\mu(x,y) \phi(x)\Big|^2-V(\phi(x)) \right).
\eea

\subsection{Breakdown of the remaining $U(1)$}

When a dark Higgs with $U(1)$ charge $+1$ sitting at site $l$ gets a VEV as $\langle\phi\rangle=\frac{1}{\sqrt{2}}\,v_\phi \ll f$, from eq.~(\ref{scalar}), we get the extra mass terms for gauge fields as follows,  
\be
\Delta {\cal L}_{\rm g, mass} = \frac{1}{2} g^2 v^2_\phi   \left(\frac{N_0}{q^l}\, {\tilde A}^0_\mu(x)+\sum_{k=1}^N  a_{lk} {\tilde A}^k_\mu(x) \right)^2. 
\ee
Then, the mass of the zero mode becomes
\bea
M^2_0\approx \frac{g^2N^2_0}{q^{2l}}\, v^2_\phi 
\eea
while
the spectrum of massive modes gets a shift as follows, 
 \be
M^2_k\approx m^2\Big(1+q^2-2q \cos \frac{k\pi}{N+1} \Big)+g^2 a^2_{lk} v^2_\phi,  \quad k=1,2,\cdots, N.
\ee
As a result, we can get the mass of the zero mode to be much smaller than $v_\phi$, due to the localization of the zero mode, while the shifts in masses of massive modes are of order $g^2 v^2_\phi/m$. For instance, for $l=N$, we get the maximum suppression of the mass of zero mode $U(1)'$ for a given $v_\phi$. On the other hand, for $l=0$, the mass of zero mode $U(1)'$ is of order the symmetry breaking scale $v_\phi$. 

From the point view of the continuum clockwork, the breaking of one of $U(1)'$ gears corresponds to a localized Higgs mechanism on the orbifold fixed point (IR brane). Therefore, we can link the suppressed mass of the zero mode by the warp factor of the metric as follows,
\bea
M_0\approx e^{-ky_c}\, g \, v_{\rm IR}
\eea
where $v_{\rm IR}$ is the VEV of the dark Higgs field localized on the IR brane, which is of order the cutoff scale on the IR brane. As a consequence, it is possible to have a light gauge boson mass hierarchically smaller than the cutoff scale of the theory.  If $v_\phi$ of order the electroweak scale is explained by the warp factor in the extra dimension, the dark gauge boson receives a smaller mass due to the localization. 
On the other hand, for $v_{\rm IR}\ll R^{-1}$, we can ignore the corrections to the KK masses due to the VEV of the brane-localized Higgs field.

\section{Examples for $U(1)$ clockwork }

We consider examples for the clockwork with anomaly-free $U(1)'$, composed of identical $N+1$ $U(1)'$, beyond the SM. The $U(1)$ clockwork can be dark $U(1)$ or  $U(1)_{B-L}$ or other anomaly-free $U(1)$'s with extra chiral fermions. In this section, we regard the $U(1)$ clockwork as the mediators of dark matter interactions, flavor-changing interactions for $B$-meson decays, and D-term SUSY breaking. 

The abelian gauge bosons of the gauged $U(1)$ clockwork are denoted as $Z^{\prime i}_\mu$ with $i=0,1,2,\cdots,N$.  After the link Higgs fields get VEVs, $N+1$ $U(1)'$ are broken down to one $U(1)'$.
We also introduce extra Higgs field to break the remaining $U(1)'$ on one of the sites. 
The setup can easily generalized to a supersymmetric case where the D-terms and superpartners of gauge bosons, i.e. gauginos, are included in the superfield Lagrangian.

\subsection{$U(1)$ clockwork as the mediator of dark matter}

We consider a Dirac fermion dark matter (DM) that is vector-like under one of $U(1)'$ gears such that it has a small or large effective gauge coupling to the zero mode of $U(1)'$, depending on the location of the $U(1)'$ gear. We discuss the implications of the localized zero mode of $U(1)'$ for DM interactions.
We also include the discussion on the case with axial vector couplings of dark matter. 
Although we don't consider a concrete anomaly-free $U(1)'$ model in this work, it might be possible to have a  fermion dark matter with axial coupling when it carries a chiral charge and mixes with heavy fermion with opposite chiral charge.

Suppose that the SM fermions $f$ with vectorial coupling $v_f$ are localized on site $j=N$ and the DM fermion $\chi$ with vectorial coupling $v_\chi$ is localized on site $j=0$.  In this case, the zero mode of $U(1)'$ may play a role as light mediator for dark matter.
In this case, the gauge couplings to mass eigenstates of gauge fields are given by
\bea
{\cal L}_{\rm Z'} &=& - v_f g_{Z'}{\bar f}(x)\gamma^\mu  f(x) \left(\frac{N_0}{q^N}\, {\tilde Z}^0_\mu(x)+\sum_{k=1}^N  a_{Nk} {\tilde Z}^k_\mu(x) \right) \nonumber \\
&&-v_\chi g_{Z'}{\bar \chi}(x)\gamma^\mu \chi(x) \left(N_0\, {\tilde Z}^0_\mu(x)+\sum_{k=1}^N  a_{0k} {\tilde Z}^k_\mu(x) \right)
\eea
where ${\tilde Z}^k_\mu$ with $k=0,1,2,\cdots,N$ are mass eigenstates of $U(1)$'s and $g_{Z'}$ is the $U(1)'$ gauge coupling.  

Moreover, as discussed in the previous section, we introduce a complex scalar field with a nonzero $U(1)'$ charge on site $j=l$ that breaks the remaining $U(1)'$ symmetry with a nonzero VEV $v_\phi$. Then, the zero mode of $U(1)'$ get mass, $M_0\approx g_{Z'} N_0 v_\phi /q^l$. 

As a consequence, the SM fermions become mili-charged under the zero mode of $U(1)'$ while the DM fermion carries a charge of order one under the same symmetry. Therefore, it is possible to have a large self-interaction of dark matter due to the sizable coupling to the zero mode of $U(1)'$, while the annihilation cross section of dark matter into the SM fermions with the zero mode mediator is suppressed.  However, the heavy gears of the $U(1)'$ clockwork have sizable couplings to both the SM fermions and the dark matter fermion, so they could be relevant for the annihilation of dark matter. 

In the limit of non-relativistic dark matter, the annihilation cross section of dark matter into a pair of SM fermions such as $\chi{\bar\chi}\rightarrow f{\bar f}$ is given by
\bea
\langle\sigma v\rangle_{\chi{\bar\chi}\rightarrow f{\bar f}}&=&\frac{v^2_f v^2_\chi g^4_{Z'}}{2\pi}\, (m^2_f+ 2m^2_\chi) \sqrt{1-\frac{m^2_f}{m^2_\chi}}\,\cdot \bigg|  \frac{N^2_0/q^N}{4m^2_\chi-M^2_0+i \Gamma_0 M_0} \nonumber \\
&&+\sum_{k=1}^N \frac{a_{0k} a_{Nk}}{4m^2_\chi-M^2_k+i\Gamma_k M_k} \bigg|^2  \label{annx}
\eea
while the annihilation cross section for $\chi{\chi}\rightarrow {\tilde Z}^0{\tilde Z}^0$ is
\bea
\langle\sigma v\rangle_{\chi{\bar\chi}\rightarrow {\tilde Z}^0 {\tilde Z}^0}&=&\frac{ N^2_0 v^4_\chi g^4_{Z'} }{4\pi}\,\frac{m^2_\chi}{(M^2_0-2m^2_\chi)^2}\,\Big(1-\frac{M^2_0}{m^2_\chi} \Big)^{3/2}.
 \label{annx2}
\eea
Then, the total annihilation cross section of dark matter is given by
$\langle\sigma v\rangle_{\rm ann}\equiv \langle\sigma v\rangle_{\chi{\bar\chi}\rightarrow f{\bar f}}+\langle\sigma v\rangle_{\chi{\bar\chi}\rightarrow {\tilde Z}^0 {\tilde Z}^0}$. 
If $\chi{\chi}\rightarrow {\tilde Z}^0{\tilde Z}^0$ is open, it dominates the dark matter annihilation. 
When dark matter is heavy enough, it also self-annihilates into a pair of heavy states, ${\tilde Z}^k$, but we don't consider this possibility in the later discussion.  

On the other hand, the DM self-scattering cross section for $\chi{\chi}\rightarrow \chi{\chi}$  is given by
\bea
\sigma_{ \chi{\chi}\rightarrow\chi{\chi} }= \frac{v^4_\chi g^4_{Z'} m^2_\chi}{8\pi}\, \left|  \frac{N^2_0}{M^2_0}+\sum_{k=1}^N \frac{a_{0k} a_{0k}}{M^2_k} \right|^2, \label{selfx}
\eea
The corresponding cross section for $\chi{\bar\chi}\rightarrow \chi{\bar\chi}$ can be similarly obtained.
Then, the effective self-scattering cross section is given by $\sigma_{\rm self}\equiv \frac{1}{4}(\sigma_{ \chi{\chi}\rightarrow\chi{\chi} }+\sigma_{ {\bar \chi}{\bar\chi}\rightarrow{\bar\chi}{\bar\chi} }+\sigma_{\chi{\bar\chi}\rightarrow \chi{\bar\chi}})$. 
Likewise, the DM-nucleon elastic scattering cross section for $\chi N\rightarrow \chi N$ is given by
\bea
\sigma_{\chi N}=\frac{\mu^2_N}{\pi A^2}\,\Big(Z f_p + (A-Z) f_n \Big)^2 
\eea
where $\mu_N\equiv m_N m_\chi/(m_N+ m_\chi)$ is the reduced mass for the DM-nucleon system, and $Z, A$ are the number of protons and nucleons, respectively, and the effective interactions between DM and protons or neutrons are given by
\bea
f_{p,n} = c_{p,n} v_\chi q^2_{Z'} \left(\frac{N^2_0/q^N}{M^2_0}+\sum_{k=1}^N \frac{a_{0k} a_{Nk}}{M^2_k} \right)  \label{nucleonint}
\eea
with $c_p\equiv 2q_u+q_d$ and $c_n\equiv q_u+2q_d$.
On the other hand, in the case of light dark matter of sub-GeV scale, the DM-electron scattering process, $\chi \,e\rightarrow \chi\,e$, becomes more important and the corresponding scattering cross section is,  for $m_e,m_\chi, M_0\gg m_\chi v_{\rm DM}$, given  by
\bea
\sigma_{\chi e}= \frac{v^2_e g^4_{Z'}\mu^2_e}{\pi}\left|  \frac{N^2_0}{M^2_0}+\sum_{k=1}^N \frac{a_{0k} a_{Nk}}{M^2_k} \right|^2  \label{elecscatt}
\eea
with $\mu_e\equiv m_e m_\chi/(m_e+ m_\chi)$ being the reduced mass for the DM-electron system.

Using eqs.~(\ref{zero})-(\ref{norm}), the DM annihilation cross section (\ref{annx}) becomes in the continuum limit
\bea
\langle\sigma v\rangle_{\chi{\bar\chi}\rightarrow f{\bar f}}&\approx &\frac{v^2_f v^2_\chi g^4_{Z'}}{2\pi}\, (m^2_f+ 2m^2_\chi) \sqrt{1-\frac{m^2_f}{m^2_\chi}}\,\times \bigg| \frac{e^{-k\pi R}}{4m^2_\chi-M^2_0+i \Gamma_0 M_0} \nonumber \\
&&+\frac{1-e^{-2k\pi R}}{k\pi R}\sum_{n=1}^N \frac{n^2 \cos(n\pi)}{ R^2M^2_n(4m^2_\chi-M^2_n+i\Gamma_n M_n)} \bigg|^2  \nonumber \\
&\approx& \frac{v^2_f v^2_\chi g^4_{Z'}}{2\pi}\, (m^2_f+ 2m^2_\chi) \sqrt{1-\frac{m^2_f}{m^2_\chi}}\,\times \left| \frac{e^{-k\pi R}}{M^2_0-4m^2_\chi}+\frac{1-e^{-2k\pi R}}{k\pi R}\sum_{n=1}^N \frac{n^2 \cos(n\pi)}{R^2 M^4_n} \right|^2  \nonumber \\
\eea
where use is made of the matching condition between the 4D and 5D gauge couplings by $g^2_{Z'}=k g^2_5/(1-e^{-2\pi k R})$ for the localized zero mode\footnote{The gauge coupling for a flat zero mode is recovered as $g^2_{Z'}\simeq g^2_5/(2\pi R)$ for $\pi k R\ll 1$. } and $M_k\gg 2m_\chi$ for heavy states with $k=1,2,\cdots, N$ is taken in the second approximation. 
Similarly, the continuum limit of the DM self-scattering cross section (\ref{selfx}) is
\bea
\sigma_{\rm \chi{\chi}\rightarrow\chi{\chi} }\approx \frac{v^4_\chi g^4_{Z'} m^2_\chi}{8\pi}\, \left|  \frac{1}{M^2_0}+\frac{1-e^{-2k\pi R}}{k\pi R}\sum_{n=1}^N \frac{n^2}{R^2 M^4_n} \right|^2.  
\eea
Likewise, the continuum limits of the DM-nucleon effective interactions (\ref{nucleonint}) are
\bea
f_{p,n} \approx c_{p,n} v_\chi g^2_{Z'}\left( \frac{e^{-k\pi R}}{M^2_0}+\frac{1-e^{-2k\pi R}}{k\pi R}\sum_{n=1}^N \frac{n^2 \cos(n\pi)}{R^2 M^4_n} \right).
\eea
Moreover, the continuum limit of the DM-electron scattering cross section (\ref{elecscatt}) is
\bea
\sigma_{\chi e}\approx  \frac{v^2_e g^4_{Z'}\mu^2_e}{\pi}\left|  \frac{e^{-k\pi R}}{M^2_0}+\frac{1-e^{-2k\pi R}}{k\pi R}\sum_{k=1}^N \frac{n^2\cos (n\pi)}{ R^2 M^4_n} \right|^2.
\eea

Then, using the sum
\bea
\sum_{n=1}^\infty\frac{\cos(nx)}{n^2+\alpha^2}&=& \frac{\pi}{2\alpha} \,\frac{\cosh\alpha (\pi-x)}{\sinh(\alpha\pi)}- \frac{1}{2\alpha^2}
\eea
and its derivative with respect to $\alpha^2$, we get the formulae for the continuum limit in the closed form, as follows,
\bea
\langle\sigma v\rangle_{\rm ann}&\approx& \frac{v^2_f v^2_\chi g^4_{Z'}}{2\pi}\, (m^2_f+ 2m^2_\chi) \sqrt{1-\frac{m^2_f}{m^2_\chi}}\,\nonumber \\
&&\quad\times  \left| \frac{e^{-k\pi R}}{M^2_0-4m^2_\chi}
+\frac{1}{4k^2}\,(1-e^{-2k\pi R})\,\bigg(\frac{\sinh(k\pi R)-k\pi R\cosh(k\pi R)}{\sinh^2(k\pi R)}\bigg) \right|^2 \nonumber \\
&&+\frac{ v^4_\chi g^4_{Z'} }{4\pi}\,\frac{m^2_\chi}{(M^2_0-2m^2_\chi)^2}\,\Big(1-\frac{M^2_0}{m^2_\chi} \Big)^{3/2},  \label{ann-cont} \\
 \sigma_{\rm self}&\approx&  \frac{v^4_\chi g^4_{Z'} m^2_\chi}{16\pi M^4_0}\left[1 +\frac{16 m_\chi^4 - 20 m_\chi^2 M_0^2 + 7 M_0^4}{(4 m_\chi^2 - M_0^2)^2}\right],   \label{self-cont} \\
f_{p,n} &\approx & c_{p,n} v_\chi g^2_{Z'}\left[\frac{e^{-k\pi R}}{M^2_0}
+\frac{1}{4k^2}\,(1-e^{-2k\pi R})\,\bigg(\frac{\sinh(k\pi R)-k\pi R\cosh(k\pi R)}{\sinh^2(k\pi R)}\bigg) \right],\\
\sigma_{\chi e}&\approx & \frac{v^2_e g^4_{Z'}\mu^2_e}{\pi}\left| \frac{e^{-k\pi R}}{M^2_0}
+\frac{1}{4k^2}\,(1-e^{-2k\pi R})\,\bigg(\frac{\sinh(k\pi R)-k\pi R\cosh(k\pi R)}{\sinh^2(k\pi R)}\bigg) \right|^2.
\eea

When dark matter has only the axial coupling $a_\chi$ and the SM fermions have only the vectorial coupling $v_f$, the DM-nucleon/electron scattering cross sections are velocity-suppressed. 
Moreover, the annihilation cross section (\ref{ann-cont}) and self-scattering cross section (\ref{self-cont}) are replaced, respectively, by
\bea
\langle\sigma v\rangle_{\rm ann}&\approx& \frac{v^2_f a^2_\chi g^4_{Z'}}{12\pi}\, (m^2_f+ 2m^2_\chi) v^2 \sqrt{1-\frac{m^2_f}{m^2_\chi}}\,\nonumber \\
&&\quad\times  \left| \frac{e^{-k\pi R}}{M^2_0-4m^2_\chi}
+\frac{1}{4k^2}\,(1-e^{-2k\pi R})\,\bigg(\frac{\sinh(k\pi R)-k\pi R\cosh(k\pi R)}{\sinh^2(k\pi R)}\bigg) \right|^2 \nonumber \\
&&+\frac{ a^4_\chi g^4_{Z'} }{4\pi}\,\frac{m^2_\chi}{(M^2_0-2m^2_\chi)^2}\,\Big(1-\frac{M^2_0}{m^2_\chi} \Big)^{3/2},  \label{aann-cont} \\
 \sigma_{\rm self}&\approx&  \frac{a^4_\chi g^4_{Z'} m^2_\chi}{16\pi M^4_0}\left[9 +\frac{48 m_\chi^4 - 36 m_\chi^2 M_0^2 + 7 M_0^4}{(4 m_\chi^2 - M_0^2)^2}\right].  \label{aself-cont}
\eea

\begin{figure}
  \begin{center}
    \includegraphics[height=0.45\textwidth]{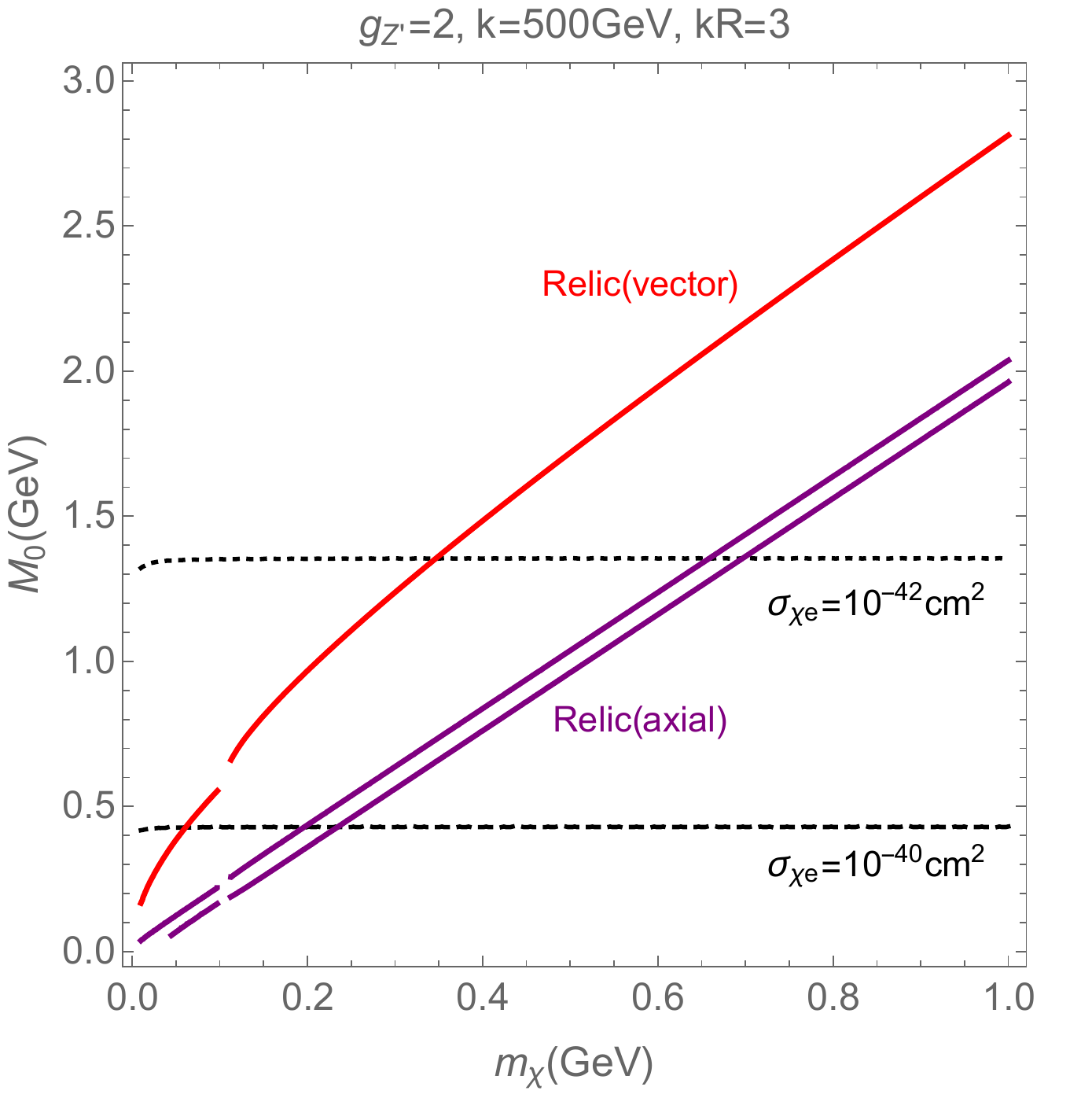}
 \includegraphics[height=0.45\textwidth]{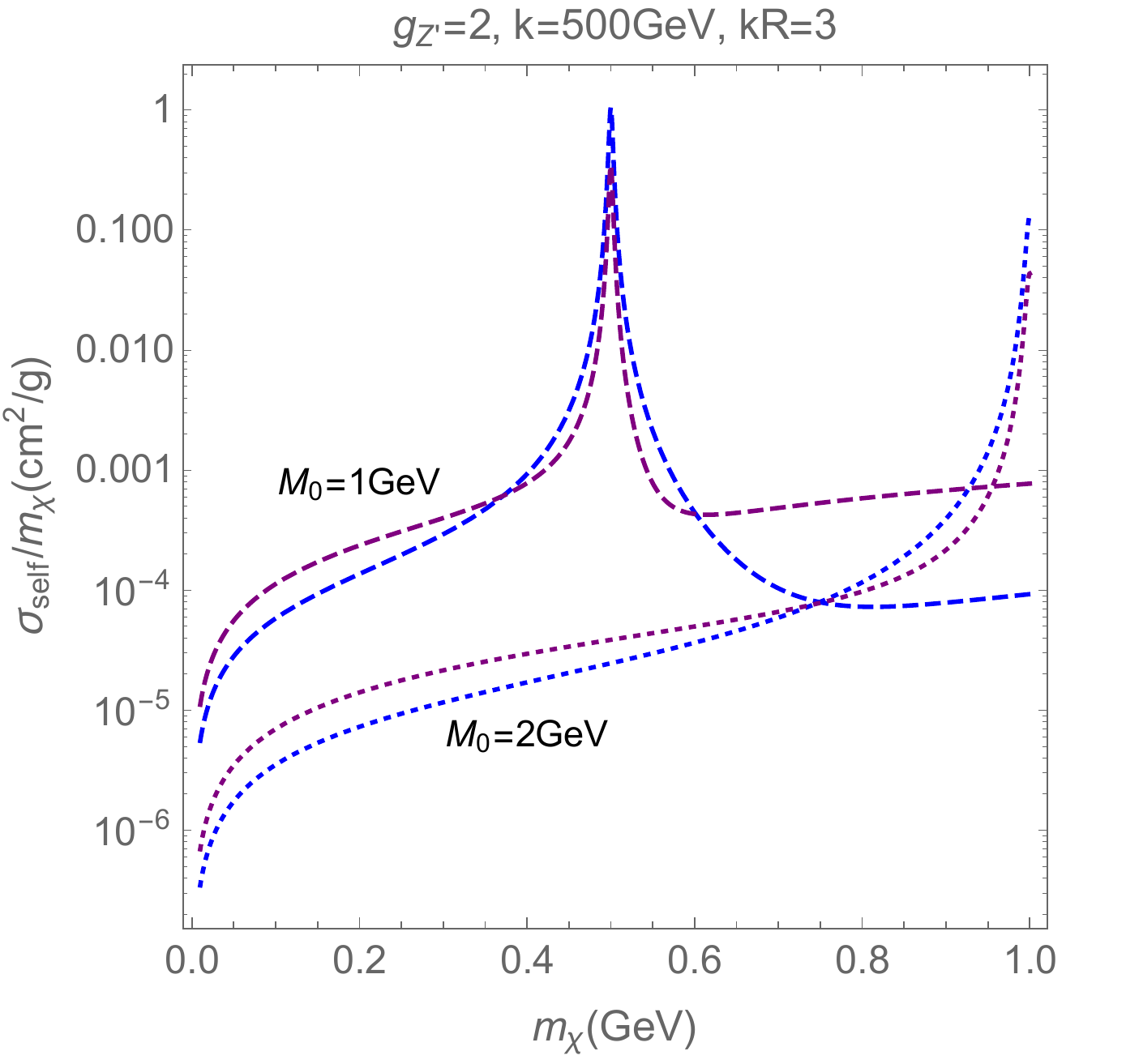}
  \end{center}
  \caption{Left: Parameter space for $m_\chi$ vs $M_0$ for light dark matter with the vectorial (axial) coupling, satisfying the relic density in red (purple) solid lines.
  Contours for the DM-electron elastic scattering cross section with $\sigma_{\chi e}=10^{-40}, 10^{-42}\,{\rm cm}^2$ are given in dashed and dotted black lines for the vectorial DM coupling only. 
  Right: DM self-scattering cross section divided by DM mass, $\sigma_{\rm self}/m_\chi$, as a function of dark matter mass. $M_0=1, 2\,{\rm GeV}$ are chosen for dashed and dotted blue (purple) lines for the vectorial (axial) DM coupling, respectively.  In both figures, we have chosen $g_{Z'}=2, k=500\,{\rm GeV}$ and $kR=3$.  }
  \label{dm1}
\end{figure}

As a result, the zero mode of $U(1)'$ can contribute to the DM self-scattering cross section significantly such that it could resolve the small-scale problems at galaxies for sub-GeV DM masses \cite{soomin} or weak-scale DM masses with Sommerfeld enhancement \cite{cirelli}. 
On the other hand, the contributions of the zero and massive modes of $U(1)'$ gears to the annihilation and DM-nucleon scattering processes are suppressed because of the localization with $e^{-k\pi R}$ in the former case or large masses of $k$ in the latter case.  We find that in the limit of $k \pi R\gg 1$, the contributions of massive modes to the DM annihilation and DM-nucleon scattering processes are highly suppressed due to the cancellations between the massive modes, whereas the contributions of them to the DM self-scattering cross section are saturated to a constant value. 
Depending on the suppression factor $1/q^N=e^{-k\pi R}$ and the mass gap $m=k$ as well as $g_{Z'}$, the interactions of $U(1)'$ clock gears could be large enough to determine the DM abundance from the freeze-out mechanism. In this case, it would be important to check the phenomenological constraints on the model from direct and indirect detection of dark matter and collider searches, etc, in a particular gauged $U(1)$ clockwork model. 
Henceforth, we discuss general aspects of the gauged $U(1)$ clockwork for dark matter physics in a model-independent way.

In the left of Fig.~\ref{dm1}, we illustrated the parameter space for $m_\chi$ vs $M_0$, in the case of light dark matter with sub-GeV scale mass. We first took into account the relic density condition in red (purple) solid line for the vectorial (axial) DM coupling and included contours of the DM-electron elastic scattering cross section in dashed and dotted black lines for $\sigma_{\chi e}=10^{-40}, 10^{-42}\,{\rm cm}^2$, respectively, for the vectorial DM coupling. Here, we have taken $v_\chi=v_f=1$, the gauge coupling for the $U(1)$ clockwork to be $g_{Z'}=2$ and the mass scale of heavy clockwork states to be $k=500\,{\rm GeV}$ and the localization parameter $kR=3$, which is equivalent to the effective coupling of the SM fermion with $ g_{Z'} e^{-k\pi R}=1.6\times 10^{-4}$.  Thus, most of the parameter space is consistent with the current limits from XENON10 \cite{xenon10}. 
In the case with $m_{Z'}>m_\chi$,  the annihilation into a pair of light fermions in the SM determines the relic density with a small effective coupling to the zero mode of $U(1)'$. On the other hand, for $m_{Z'}<m_\chi$, $\chi{\bar \chi}\rightarrow {\tilde Z}^0{\tilde Z}^0$ is dominant but thermal relics with sub-GeV scale mass could be obtained only for a small $g_{Z'}$.

Since the zero mode of $U(1)'$ decays into a pair of dark matter in most of the parameter space,  the bounds on the invisible decay mode of $Z'$ such as BaBar \cite{babar} and Belle2 in prospect \cite{belle2} as well as the beam dump experiment NA64 at CERN SPS \cite{beamdump} could be also applied. The model is consistent with the bounds from direct detection as well as the current limit from BaBar, which is at the level of $g_{Z'} e^{-k\pi R}=3\times 10^{-4}$ \cite{babar}.  

In the right of Fig.~\ref{dm1}, we also showed the DM self-scattering cross section divided by DM mass as a function of $m_\chi$ for $M_0=1,2 \,{\rm GeV}$ in blue (purple) dashed and dotted lines for the vectorial (axial) DM coupling. The values of the other parameters have been chosen the same as in the left of Fig.~\ref{dm1}. It turns out that at the resonance with $M_0\approx 2m_\chi$, the self-scattering cross section reaches the peak value close to $\sigma_{\rm self}/m_\chi=0.1-1\,{\rm cm^2/g}$ as required to solve small-scale problems at galaxies \cite{small-scale1,small-scale2}.

We remark that thermal relics with s-wave annihilation are strongly constrained by Cosmic Microwave Background (CMB) at recombination. For instance, the dark matter with mass $\lesssim 44\,{\rm GeV}$ annihilating into $e^+ e^-$ with $100\%$ is excluded by the Planck data at $95\%$ C.L.\cite{planck}.
The light dark matter with s-wave annihilation can be constrained by gamma-ray searches too \cite{gamma-rays}. 
However, if the annihilation cross section of dark matter into light fermions in the SM is p-wave suppressed, there is no such constraint on thermal dark matter.  In our model, this is the case when the $U(1)$ clockwork has an axial coupling to dark matter and a vectorial coupling to the SM fermions.  Then, the discussion on the relic density condition changes little, even if a relatively larger $g_{Z'}$ or a smaller $M_0$ is needed. Moreover, the DM-electron scattering cross section is velocity-suppressed.
The more general cases will be discussed in more detail elsewhere.

\begin{figure}
  \begin{center}
    \includegraphics[height=0.45\textwidth]{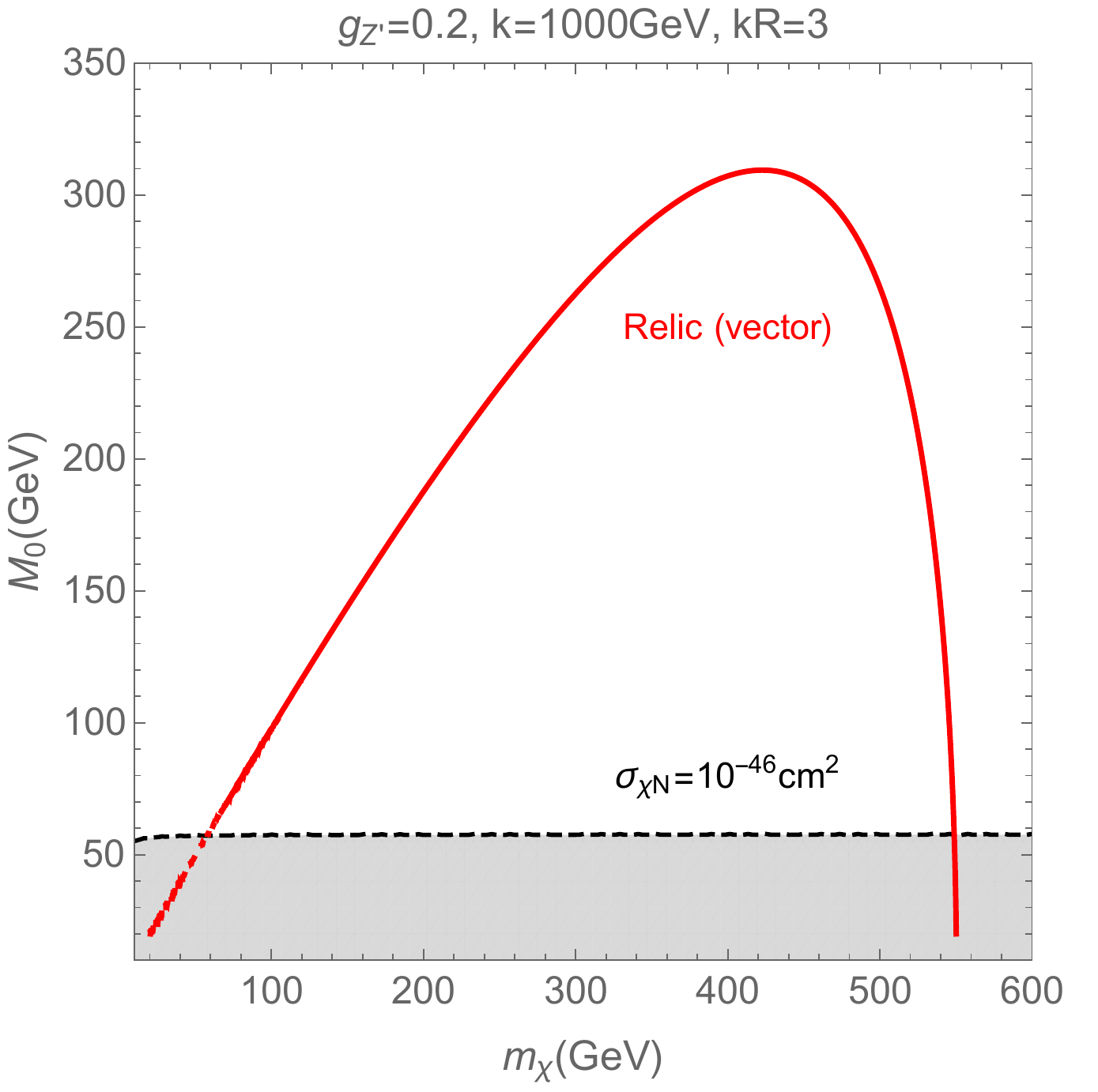}
  \end{center}
  \caption{Parameter space for $m_\chi$ vs $M_0$ for weak-scale dark matter with vectorial coupling,  satisfying the relic density in red solid line. The gray region is where the DM-nucleon elastic scattering cross section is greater than $\sigma_{\chi N}=10^{-46}\,{\rm cm}^2$. We have chosen $g_{Z'}=0.2, k=1000\,{\rm GeV}$ and $kR=3$. }
  \label{dm2}
\end{figure}

In Fig.~\ref{dm2}, we also depicted the parameter space for $m_\chi$ vs $M_0$, in the case of weak-scale dark matter with vectorial coupling. We took into account the relic density condition in red solid line and showed the region in gray where the DM-nucleon elastic scattering cross section  is greater than $\sigma_{\chi N}=10^{-46}\,{\rm cm}^2$.   Thus, most of the parameter space for weak-scale dark matter is consistent with the current limits from direct detection experiments \cite{lux,panda2,xenon1t}. Here, we have chosen $v_\chi=v_f=1$, $g_{Z'}=0.2$, $k=1000\,{\rm GeV}$ and $kR=3$, which is equivalent to the effective coupling of the SM fermion with $ g_{Z'} e^{-k\pi R}=1.6\times 10^{-5}$. In this case, the DM relic density is determined dominantly by  $\chi{\bar \chi}\rightarrow {\tilde Z}^0{\tilde Z}^0$ and it is almost independent of  whether the DM coupling is vectorial or axial. The zero mode of $U(1)'$ could not decay into a pair of dark matter in the parameter space where the relic density is explained, but instead it could decay into a pair of SM fermions. But, as ${\tilde Z}^0$ couples weakly to the SM due to a small coupling or a velocity suppression for the axial DM coupling, we can evade the bounds from Planck data at recombination and gamma-rays. 
Nonetheless, although not specified, the decay products of ${\tilde Z}^0$ might be a smoking gun signal for gamma-ray searches, depending on the decay modes of ${\tilde Z}^0$. 

If the 2-to-2 annihilation cross section is too suppressed, we can instead use the 3-to-2 processes such as $\chi {\bar\chi} \chi \rightarrow \chi {\tilde Z}^0_\mu$ to produce the dark matter from the generalized thermal freeze-out \cite{cline,vsimp}.  Even in this case, a minimum amount of the interaction between dark matter and the SM particles is needed for kinetic equilibrium of dark matter \cite{cline,vsimp}.

\subsection{$U(1)$ clockwork for $B$-meson decays}

We consider family-dependent couplings for quarks and leptons in the clockwork theory of the local $B-L$ symmetry. 
Suppose that the third family couples to the $U(1)_{B-L}$ gear at site $j=0$ while the first and second families couple to different gears at site $j=l$. 	In order to cancel the gauge anomalies of the $U(1)_{B-L}$ gears, we need to introduce one right-handed neutrino per generation. 
Then, it is possible to realize family-dependent couplings under the zero mode of $U(1)_{B-L}$ gears. 
Here, we assume that the remaining $U(1)_{B-L}$ is broken by dark Higgs fields with nonzero $B-L$ charge at least at site $j=0,  l$  for giving masses to right-handed neutrinos. 

The Lagrangian for the couplings of the $U(1)_{B-L}$ clockwork to the SM fermions is given by
\bea
{\cal L}_{B-L} &=& -g_{B-L} \bigg( \sum_{f=f_3}{\bar f} \gamma^\mu Q_{B-L} A^0_\mu f + \sum_{f=f_1, f_2} {\bar f} \gamma^\mu Q_{B-L} A^{l}_\mu f  \bigg). 
\eea
Then, the gauge interactions to the zero mode of $U(1)_{B-L}$ become
\bea
{\cal L}_{B-L} &=& -N_0 g_{B-L} {\tilde A}^0_\mu \bigg( \sum_{f=f_3}{\bar f} \gamma^\mu Q_{B-L}  f + \sum_{f=f_1,f_2} \frac{1}{q^{l}}\, {\bar f} \gamma^\mu Q_{B-L}  f   \bigg) \\
&=& -N_0 g_{B-L} {\tilde A}^0_\mu \Big(\frac{1}{3}({\bar t}\gamma^\mu t+{\bar b}\gamma^\mu b)-{\bar \tau}\gamma^\mu \tau -{\bar\nu}_\tau\gamma^\mu \nu_\tau+\cdots \Big).
\eea
Therefore, for $q^{l}\gg 1$, we get the effective couplings of the first and second families to the zero mode of $U(1)_{B-L}$ to be suppressed. As a result, the $U(1)_{B-L}$ clockwork realizes the generation-dependent $B-L$, that is, $U(1)_{B_3-L_3}$  \cite{yanagida,ligong}, that has been proposed to explain the $B$-meson anomalies observed recently at LHCb \footnote{One might think that flavor-dependent $B-L$ coupling to leptons could explain the $B$-meson anomalies, but this is not true because of a wrong sign of the effective operator, $({\bar b}\gamma^\mu P_L s)({\bar\mu}\gamma_\mu \mu)$, as compared to the global fit results \cite{crivellin,ligong}.}. 

In order to explain the $B$-meson anomalies in $R_K$ and $R_{K^*}$ \cite{Bmeson} by the modified effective operator for  ${\bar b}\rightarrow {\bar s}\mu^+ \mu^-$, we need to introduce flavor violating interactions for bottom quark and muon couplings to the extra gauge boson ${\tilde A}^0_\mu$. The flavor violating interactions of ${\tilde A}^0_\mu$ to bottom quark can be induced by the quark mixings in the CKM matrix \cite{ligong}. Violation of the lepton flavor universality can be achieved by the lepton flavor mixing \cite{yanagida} or the mixing between  $B_3-L_3$ and $L_\mu-L_\tau$ \cite{ligong}. 

We focus on the case that there is a mixing between the effective $B_3-L_3$ (i.e. the zero mode of the $U(1)_{B-L}$ clockwork) and $L_\mu-L_\tau$ \cite{ligong}, namely, the $U(1)'$ gauge symmetry at low energy is given by $x(B_3-L_3)+y (L_\mu-L_\tau)$. In this case, after integrating out $Z'$ with mass $m_Z'$ and gauge coupling $g_{Z'}$, we get the effective Hamiltonian for ${\bar b}\rightarrow {\bar s}\mu^+\mu^-$ \cite{ligong}, as follows,
\bea
\Delta {\cal H}_{{\rm eff},{\bar b}\rightarrow {\bar s}\mu^+ \mu^-} = -\frac{4G_F}{\sqrt{2}}  \,V^*_{ts} V_{tb}\,\frac{\alpha_{em}}{4\pi}\, C^{\mu,{\rm NP}}_9 {\cal O}^\mu_9
\eea
with $ {\cal O}^\mu_9 \equiv ({\bar s}\gamma^\mu P_L b) ({\bar \mu}\gamma_\mu \mu)$ and $\alpha_{\rm em}$ being the electromagnetic coupling, we obtain a new physics contribution to the Wilson coefficient as follows,
\bea
C^{\mu, {\rm NP}}_9= -\frac{8 xy \pi^2\alpha_{Z'}}{3\alpha_{\rm em}}\, \Big(\frac{v}{m_{Z'}}\Big)^2
\eea
with $\alpha_{Z'}\equiv g^2_{Z'}/(4\pi)$. From  the best-fit value, $C^{\mu, {\rm NP}}_9=-1.10$ \cite{crivellin}, (while taking $[-1.27,-0.92]$ and $[-1.43,-0.74]$ within $1\sigma$ and $2\sigma$ errors), to explain the $B$-meson anomalies, we need \cite{ligong}
\be
m_{Z'}=\Big(xy\, \frac{\alpha_{Z'}}{\alpha_{\rm em}} \Big)^{1/2}\, 1.2\,{\rm TeV}.  
\ee
Various constraints on the $Z'$ interactions, coming from dimuon resonance searches, other meson decays and mixing, tau lepton decays and neutrino scattering, have been studied in detail in Ref.~\cite{ligong}, leading to the conclusion that $x g_{Z'}\lesssim 0.05$ for $y g_{Z'}\sim 1$ and $m_{Z'}\lesssim 1\,{\rm TeV}$. 

We remark the Yukawa couplings for quarks and leptons in the case of $U(1)_{B-L}$ clockwork. 
Although the SM fermions couple to the $U(1)_{B-L}$ gears in a family-dependent way, their Yukawa couplings are invariant under the $U(1)_{B-L}$ gears. But, the hierarchy of fermion masses and mixings depend on the profile of the zero mode of the SM Higgs doublet.  The concrete discussion on the case that the Yukawa interactions are restricted only by the gauge symmetry has been studied \cite{ligong}.

\subsection{D-term SUSY breaking}

In this subsection, we discuss another example of utilizing the $U(1)$ clockwork for mediating the SUSY breaking to the visible sector. 

The hidden sector SUSY breaking in supergravity leads to F-term of order $\langle F \rangle={\cal O}(m_{3/2} M_P)$ and D-term of order $\langle D\rangle={\cal O}(m^2_{3/2})$ where $m_{3/2}$ is the gravitino mass \cite{Dterm}. Thus, the D-term SUSY breaking can be important when the gravitino is quite heavy. 
However, the soft masses in the visible sector depend on the mediation mechanisms of SUSY breaking. 
In particular, when the gravity mediation of F-term SUSY breaking is suppressed by a sequestering mechanism \cite{sequestering}, for instance, when the hidden and visible sectors are localized at different fixed points in the extra dimension, the D-term SUSY breaking could be dominant \cite{U1R}. 

We consider a clockwork Lagrangian for the hidden D-terms of $N+1$ $U(1)$ vector superfields with $D_j$ ($j=0,1,\cdots,N-1$), as follows,
\bea
{\cal L}_D=\sum_{j=0}^{N-1} \frac{1}{2} (D_j- qD_{j+1})^2.
\eea
We note that the above component Lagrangian can be derived from the superfield Lagrangian, ${\cal L}_D=\frac{1}{32g^2}\int d^2\theta \sum_{j=0}^{N-1} (W_j-qW_{j+1} )_\alpha(W_j-q W_{j+1})^\alpha+{\rm h.c.}$, where $W_j$ are the field strength superfields. 
Then, the zero mode of the D-term is taken along the direction satisfying the relations between clock gears, $D_N=\frac{1}{q}\, D_{N-1}=\frac{1}{q^2}\, D_{N-2}=\cdots=\frac{1}{q^{N-1}}\, D_1=\frac{1}{q^N} D_0$. 
Therefore, for a large D-term SUSY breaking on site $j=0$, the effective D-term on site $j=N$ is suppressed by a factor of $1/q^N$ for $q>1$. 

Now we add the source of the D-term SUSY breaking on site $j=0$ in the following form,
\bea
\Delta {\cal L}_D= \delta_0 \bigg(\frac{1}{2} D^2_0 -\xi_0 D_0 \bigg)
\eea
where $\delta_0$ is an arbitrary constant and $\xi_0$ is the Fayet-Iliopoulos D-term.  
Here, we note that the additional term  corresponds to  the superfield Lagrangian, $\frac{1}{32g^2}\delta \Big(\int d^2\theta (W_0)_\alpha (W_0)^\alpha+{\rm h.c.}-\int d^4 \theta\, \xi V_0\Big)$. 
Then, the equation of motion for D-terms leads to $D_N=\frac{1}{q^N}\, D_0=\frac{\xi}{q^N}$, resulting in the suppressed D-term SUSY breaking on site $j=N$. 
Now including a complex scalar field $\phi_N$ with nonzero charge $Q_{\phi_N}$ under $U(1)_N$ has a D-term coupling as follows,
\bea
{\cal L}_{D,j=N}= \frac{1}{2} q^2 \Big( D_N+  \frac{g}{q}\, Q_{\phi_N} \phi^\dagger_N \phi_N \Big)^2,
\eea
we obtain the soft scalar mass for $\phi_N$ on site $j=N$ as follows,
\bea
m^2_{\phi_N}= q g Q_{\phi_N} D_N = \frac{g Q_{\phi_N} \xi_0}{q^{N-1}}\,.  
\eea
Therefore, if the scalar superpartners of the SM fermions carry nonzero $U(1)_N$ charges on site $j=N$, the D-term clockwork mechanism can provide a small soft SUSY breaking in the visible sector. 
We note that when the D-term breaking can be introduced on site $j=N$ by $\Delta {\cal L}_D= \delta_N \bigg(\frac{1}{2} D^2_N -\xi_N D_N \bigg)$, a hierarchically small D-term can be induced on site $j=0$ by $D_0=q^N D_N$ for $q<1$.
This generalizes the previous discussion on a generation of small mass scales through a small gauge kinetic mixing in Ref.~\cite{wang}.

\section{Conclusions}

We have studied the gauged $U(1)$ clockwork theory with a product of multiple $U(1)$'s where the gauge symmetries are broken down to one $U(1)$ by the Higgs mechanism. In the continuum limit, the $U(1)$  clockwork theory corresponds to a massless gauged $U(1)$ theory in five dimensions with linear dilaton background.  We have introduced interactions of external matter fields to $U(1)$ clock gears for generating a hierarchy of couplings to the zero mode of $U(1)$ clock gears. 
We discussed the consequences of our general discussion with some examples, focusing on the mediators of dark matter interactions, flavor-changing interactions for $B$-meson decays and briefly sketching the case with D-term SUSY breaking.

\section*{Acknowledgments}

I would like to thank Soo-Min Choi and Yoo-Jin Kang for checking some formulas and discussion. 
The work is supported in part by Basic Science Research Program through the National Research Foundation of Korea (NRF) funded by the Ministry of Education, Science and Technology (NRF-2016R1A2B4008759). 

%\def\theequation{A.\arabic{equation}}

%\setcounter{equation}{0}

%\vskip0.8cm
%\noindent
%{\Large \bf Appendix A: } 
%\vskip0.4cm
%\noindent

\end{document}